# New methods of removing debris and high-throughput counting of cyst nematode eggs extracted from field soil

**Upender Kalwa**[1☯], **Christopher Legner**[1☯], **Elizabeth Wlezien**[2], **Gregory Tylka**[2], **Santosh Pandey**[1]*

**1** Department of Electrical and Computer Engineering, Iowa State University, Ames, Iowa, United States of America, **2** Department of Plant Pathology and Microbiology, Iowa State University, Ames, Iowa, United States of America

☯ These authors contributed equally to this work.
* pandey@iastate.edu

Check for updates

OPEN ACCESS

**Citation:** Kalwa U, Legner C, Wlezien E, Tylka G, Pandey S (2019) New methods of removing debris and high-throughput counting of cyst nematode eggs extracted from field soil. PLoS ONE 14(10): e0223386. https://doi.org/10.1371/journal.pone.0223386

**Editor:** Jie Zhang, Newcastle University, UNITED KINGDOM

**Received:** June 6, 2019

**Accepted:** September 19, 2019

**Published:** October 15, 2019

**Data Availability Statement:** All software files are available through gitHub. (https://github.com/ukalwa/nematode_egg_counting).

**Funding:** This work is partially supported by the U. S. National Science Foundation [NSF IDBR-1556370] to S. P. and G. T. and the Defense Threat Reduction Agency [HDTRA1-15-1-0053] to S. P.

**Competing interests:** The authors have declared that no competing interests exist.

## Abstract

The soybean cyst nematode (SCN), *Heterodera glycines*, is the most damaging pathogen of soybeans in the United States. To assess the severity of nematode infestations in the field, SCN egg population densities are determined. Cysts (dead females) of the nematode must be extracted from soil samples and then ground to extract the eggs within. Sucrose centrifugation commonly is used to separate debris from suspensions of extracted nematode eggs. We present a method using OptiPrep as a density gradient medium with improved separation and recovery of extracted eggs compared to the sucrose centrifugation technique. Also, computerized methods were developed to automate the identification and counting of nematode eggs from the processed samples. In one approach, a high-resolution scanner was used to take static images of extracted eggs and debris on filter papers, and a deep learning network was trained to identify and count the eggs among the debris. In the second approach, a lensless imaging setup was developed using off-the-shelf components, and the processed egg samples were passed through a microfluidic flow chip made from double-sided adhesive tape. Holographic videos were recorded of the passing eggs and debris, and the videos were reconstructed and processed by custom software program to obtain egg counts. The performance of the software programs for egg counting was characterized with SCN-infested soil collected from two farms, and the results using these methods were compared with those obtained through manual counting.

## Introduction

Many nematodes (microscopic roundworms) are soil-dwelling plant parasites that infect the roots of plants and cause billions of dollars of crop loss worldwide on an annual basis. *Heterodera glycines*, the soybean cyst nematode (SCN), is the most damaging soybean pathogen in the United States and Canada, causing hundreds of millions of dollars in crop loss annually [1]. The SCN infects the roots of soybeans and siphons nutrients from the plants, leading to stunted growth and reduced crop yields. The amount of damage and yield loss caused by the nematode is related to several factors, including the number of eggs (egg population density) in the soil. Knowing the SCN population density in the soil can be useful in guiding the use of management strategies and assessing the success of management efforts in SCN-infested fields.

To determine the population density of SCN in a field, one or more multiple-core soil samples are collected. On a sample-by-sample basis, the cysts (egg-filled dead SCN females) are extracted from samples, and the eggs are extracted from the cysts, then counted using a microscope. Two methods commonly used to extract nematode cysts from the soil are (i) wet sieving and decanting, and (ii) elutriation. Wet sieving and decanting [2] involve suspending soil in water, agitating the suspension, allowing the heavier soil particles to settle to the bottom of the container, and pouring the suspension through two sieves. The top sieve (usually with 850-μm-diameter pores) will capture root fragments and other debris, which will be discarded, and the bottom sieve (usually with 250-μm-diameter pores) will capture nematode cysts and cyst-sized debris. Smaller debris (<250 μm in diameter) will pass through both sieves and be discarded. For the elutriation method, soil is suspended in a column or cone of upward-flowing water. Heavy soil particles remain near the bottom of the flowing water suspension, while cysts and less dense objects float and pour out of the top of the column or cone to be captured on sieves [3] as with the wet-sieving and decanting method.

Currently, two methods commonly used for extracting eggs from nematode cysts are (i) by grinding the cysts in a plastic or glass tube with a stainless-steel pestle [4] or Teflon tissue homogenizer and (ii) by grinding the cysts on a 250-μm-pore sieve with a rubber stopper [5]. Both methods result in the capture of eggs and similarly sized debris on 25-μm-pore sieves. Accurately and efficiently counting the eggs to determine the population density can be difficult and inefficient when considerable amounts of debris are recovered with the eggs. Separating the eggs from the debris often is necessary to make counting possible. Also, it is desirable to separate and discard debris from eggs in suspension when the eggs are to be used in laboratory experiments, such as when assessing development and studying hatching of nematode juveniles from the eggs [6–8].

Sucrose centrifugation [9] is a method commonly used to separate plant-parasitic nematode juveniles and eggs in suspension from debris. Although sucrose centrifugation is inexpensive and easy, it may not be the most efficient method and it may have adverse effects on the nematodes. Exposing nematode eggs and juveniles to sucrose solutions, with a high osmotic potential, may harm the eggs and juveniles if exposure to the sucrose is prolonged. Also, if the nematodes are not thoroughly rinsed with water following centrifugation in sucrose solution, sucrose residues may remain and promote bacterial and fungal growth on the eggs and juveniles. Deng et al. [10] suggested an alternative density gradient method to sucrose centrifugation using an iodixanol solution, OptiPrep™ (also known as Visipague™ in medical uses). Testing the effects of a single concentration of OptiPrep™, they found that the efficiency of extraction of the reniform nematode (*Rotylenchulus reniformis*) and post-extraction mobility of the recovered nematodes were both 100% greater than when the sucrose centrifugation method was used.

Following the extraction and cleaning of egg samples, the eggs are counted to determine the population density of nematodes. Counting is completed manually through microscopic observation, requiring trained personnel. The process is time intensive, laborious, and prone to human error. It would be ideal if this portion of the process could be automated to address these shortcomings. Furthermore, automated egg counting could reduce labor costs and processing fees compared to manual counting methods where the expense increases with sample quantity,

The objectives of this work were: 1) to develop a method of purifying eggs from suspension with debris and 2) to develop new methods to automate counting of eggs once extracted from cysts and purified. In the first counting approach, a high-resolution scanner takes images of the processed sample (i.e. stained eggs with debris) dispersed on filter paper that was then run through deep learning algorithms to automatically identify and count the eggs. In a second counting approach, a benchtop, lensless imaging setup takes real-time, holographic videos of the processed sample passing through a flow chip and is analyzed with a custom software program to determine the egg count. We performed detailed characterization of the new methods, while attempting to minimize the number of manual steps.

## Materials and methods

### Sample preparation

Soil samples were collected from two fields in Muscatine and Story County in Iowa. The fields were located on research farms owned and operated by Iowa State University. No specific permissions were required to obtain the soil because the purpose of the farms and their fields is to support crop research, and collection of soil samples is a routine part of conducting such research. The field studies did not involve endangered or protected species. The sample preparation involved two steps: egg extraction and egg staining.

To extract the eggs, each soil sample was poured in a bucket filled with water (approximately 2 liters), mixed thoroughly, and allowed to settle. The soil suspension then was poured through a 20-cm-diameter sieve with 850-μm-diameter pores above a 20-cm-diameter sieve with 250-μm-diameter pores [2]. The debris and egg-filled cysts (dead nematode females) were collected on the 250-μm-pore sieve and then transferred to a 3.7-cm-diameter, 250-μm-pore sieve and were crushed using a motorized rubber stopper to release the eggs [5]. These eggs (along with similar sized debris) were collected on a 15-cm-diameter sieve with 37-μm-diameter pores and then transferred into a microwavable container [11].

For staining the eggs, a stain solution was prepared by adding 3.5 g acid fuchsin (F8129, Sigma Aldrich) and 250 mL glacial acetic acid (ARK2183, Sigma Aldrich) to 750 mL of distilled water and stirred well [4]. One drop of the stain and four to five drops of 1 M HCl solution were added to each plastic beaker containing the eggs. The beakers and their contents were heated in a microwave for 15 seconds to stain the eggs [12].

### Sample cleaning method using OptiPrep™-based centrifugation

Fig 1A is a schematic representation of the protocol for sample cleaning using OptiPrep™ as the density gradient medium. The protocol consists of two steps: centrifugation and separation. During the centrifugation step, 3 mL of the density gradient medium (OptiPrep™) at a specific volume percentage was put in a 15 mL centrifuge tube (Corning™). Thereafter, 5 mL of the stained egg sample was pipetted on top of the OptiPrep™ solution, forming an emulsion interface layer between the solutions of different densities as shown in Fig 1A(i.). The combined sample was centrifuged at 840 G for 2 minutes. Particles with higher density than the gradient solution passed through the emulsion layer and were deposited in a pellet at the bottom. Particles with comparable density to the gradient solution were concentrated in the interface layer, while particles having lower density floated in the top layer of the solution. The contents of the centrifuge tube were categorized into three layers: top (4 mL), interface (2 mL), and bottom (2 mL). In the separation step, each of the three layers were pipetted into three different 15 mL test tubes and diluted to 11 mL with water, as shown in Fig 1A(ii.).

### Scanner-based egg counting method

**Hardware and software components.** Fig 1B illustrates the scanner-based egg counting method. Here the three distinct layers of processed sample (top, interface, and bottom) obtained after centrifugation in OptiPrep™ were poured on separate circular filter papers (Grade 41, Whatman™, 90 mm diameter) and allowed to air dry at room temperature (approximately 22˚C). The filter papers were placed on an A4-sized (21.6 × 27.9 cm) transparency sheet face-down and placed on a flatbed scanner (Epson Perfection v750 Pro) connected to a desktop computer. Using the scanner software (Epson Scan software, Version 3.921), each piece of filter paper was selected with a bounding box and scanned as shown in Fig 1B. The following settings were chosen: reflective scanning mode, 4800 dpi, and 24-bit color. All other imaging parameters were set to their default values. The scanned images were saved as JPEG files to the computer hard drive.

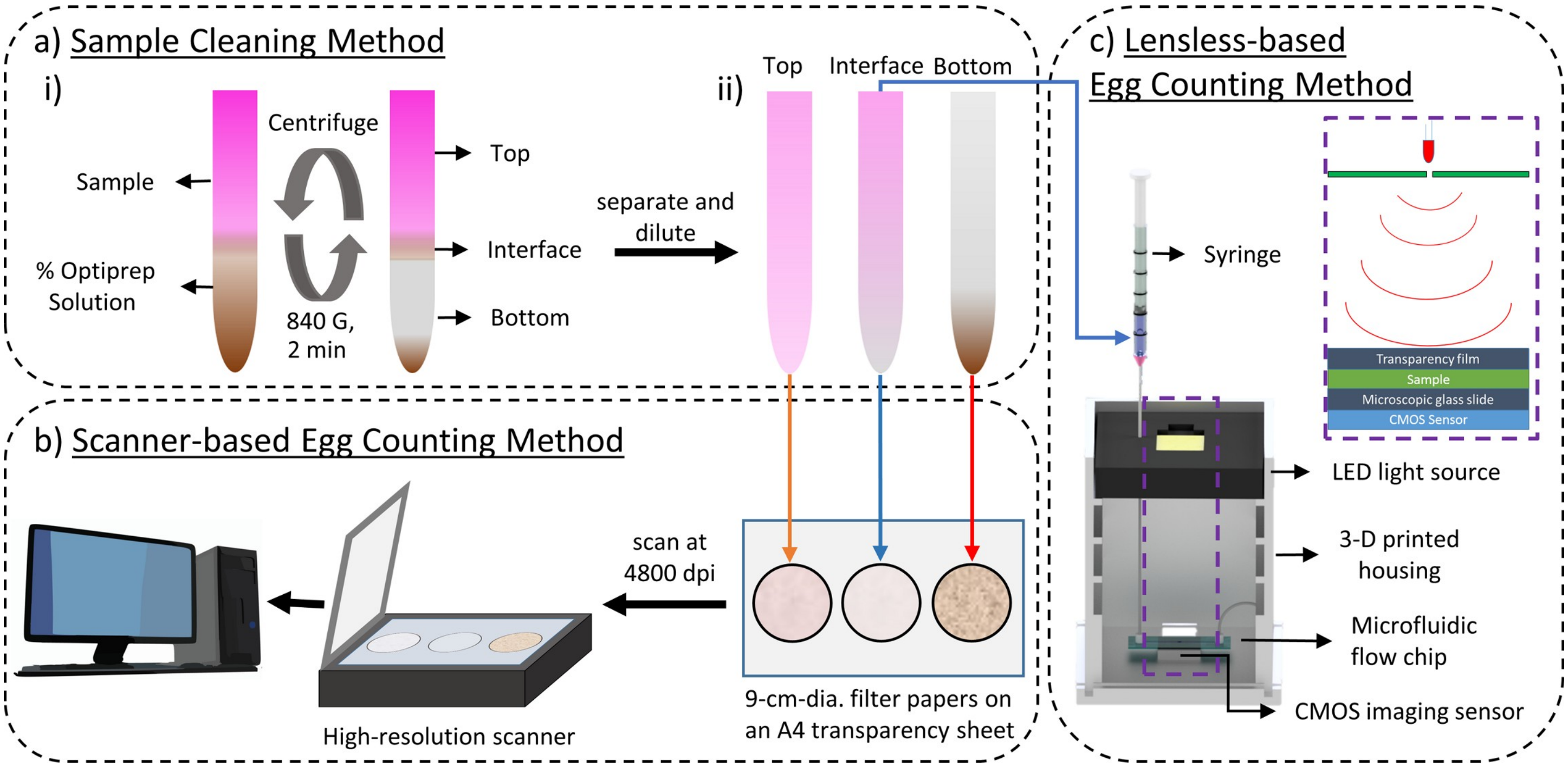

**Fig 1. Overview of the new methods of sample cleaning and egg counting.** a) Sample cleaning used a density gradient medium (OptiPrep™) and consisted of two steps: centrifugation and separation. i) An illustration of the density gradient centrifugation is shown where the three distinct layers (top, interface, and bottom) were visible after the centrifugation step. ii) The three layers were separated and diluted with water to 11 mL. b) A scanner-based egg counting method is shown where three solutions were dispersed on separate filter papers, allowed to air dry, and then placed face-down on an A4-sized (21.6 × 27.9 cm) transparency sheet. The filter papers were scanned at 4800 dpi and analyzed by a deep learning network model to obtain a count of the eggs. c) The lensless setup consisted of a light-emitting diode (LED) to illuminate the processed sample flowing through a microfluidic flow chip and a CMOS image sensor to record videos of the eggs in real-time. A custom software program, written in Python, analyzed the videos and produced the egg count.

**Data collection and pre-processing.** Each scanned image of the filter paper (17759 × 17759 pixels) was split into 4900 patch images (256 × 256 pixels). The colorspace of each patch image was converted from RGB (Red, Green and Blue) to HSV (Hue, Saturation, and Value) to distinguish the different colors because the HSV colorspace separated the color information (chroma) from the image intensity (luma). A range of HSV values was selected to identify objects of a similar color to the stained eggs within the patch image. Then by applying thresholds on the object's physical dimensions (i.e. width, height, shape, and area), all of the eggs were detected in the patch image. A label image with the eggs was created and stored along with the patch image on the hard drive. The above process was repeated for all the patch images, and subsequently used for training and testing the deep learning network. The patch and label images were resized to 128 × 128 pixels to reduce the number of training parameters and model size. The pixels were normalized to ensure that all of the features were given equal importance. The total data set consisted of 60 filter paper images from 20 different soil samples, yielding 294,000 patch images, which were randomly divided into training and test data sets (80:20 split) to be used to develop the deep learning model.

**Deep learning network architecture.** To automatically learn the features related to the SCN eggs from patch and label images, we have employed a convolutional autoencoder network —a specific type of autoencoder network that used convolutional layers [13,14]. In general, a 'convolutional network' is composed of four layers: convolution, activation, pooling, and drop-out layers. The convolution layer had neurons with weights and biases, which were updated after every iteration by a backpropagation algorithm in the training process. The activation layer consisted of a non-linearity function and performed mathematical operations on the input. An example activation function is the ReLU (Rectified Linear Unit) and has been shown to greatly accelerate the training process [13]. The pooling layer performed non-linear down-sampling on the input image by extracting the maximum or average of all the non-overlapping sub-regions in the image. This layer reduced the number of training parameters and memory footprint of the network. The dropout layer selected a random set of neurons determined by a percentage probability and set their inputs to zero making them unusable in the decision-making process of the network. Besides the convolutional network, the 'autoencoder neural network' presented an unsupervised learning platform to provide an approximate mapping of the inputs and outputs. It consisted of the 'encoder' and 'decoder' paths. The 'encoder' path compressed the input information by downsampling and learned important features, while the 'decoder' path reconstructed an approximate higher dimensional output utilizing upsampling operations.

A schematic of our network architecture is shown in Fig 2, which is a modified version of the U-Net convolutional autoencoder model [15]. We added dropout layers to prevent overfitting, replaced unpadded convolutions with padded convolutions to avoid cropping operations, and replaced the soft-max activation with sigmoid activation. We reduced the feature maps to half of the original size, which decreased the number of parameters.

As shown in Fig 2, the deep learning network took 3-channel RGB patch images as inputs and produced 1-channel grayscale label images as outputs. The network consisted of nine layers; each layer had two convolutional layers and a max pooling layer. In the 'encoder' path, each step performed repeated convolutions with a 3×3 kernel followed by ReLU activation function and a 2×2 max pooling operation. In addition, the fourth and fifth layers contained dropout layers with probabilities of 0.4 and 0.3, respectively. The filters were doubled as the downsampling operation was performed to move to the next layer. The number of filters started from 32 in the first layer and extended up to 512 in the fifth layer. Thereafter, the 'decoder' path was initialized wherein, at each step, an upsampling operation and 2×2 convolution was performed and followed by concatenation of the corresponding feature map from the 'encoder' path. Then, similar to the step in 'encoder' path, repeated convolutions with a 3×3 kernel followed by ReLU activation was performed. A final 1×1 convolution and sigmoid activation was performed to map the label image. Details of our deep learning model are listed in S1 Appendix.

**Training and testing the deep learning network model.** The patch images and their corresponding label images were used to train the network with Adaptive Moment Estimation (ADAM), which is a type of stochastic optimization of cross entropy loss [16]. In total, the network had 7,760,097 trainable parameters and was processed using an NVIDIA™ Geforce GTX 1070 Ti graphics processing unit (GPU) with 8 gigabyte memory and 2432 CUDA cores. The batch size was set to 32 and the training was performed for 100 epochs. The model implementation was written in Python using Keras [17] and Tensorflow [18] libraries.

The test dataset comprising the patch images was run through the above trained model and the corresponding grayscale images were generated. A thresholding technique yielded binary images which were passed through a blob labelling algorithm to get the egg counts in the patch images [19].

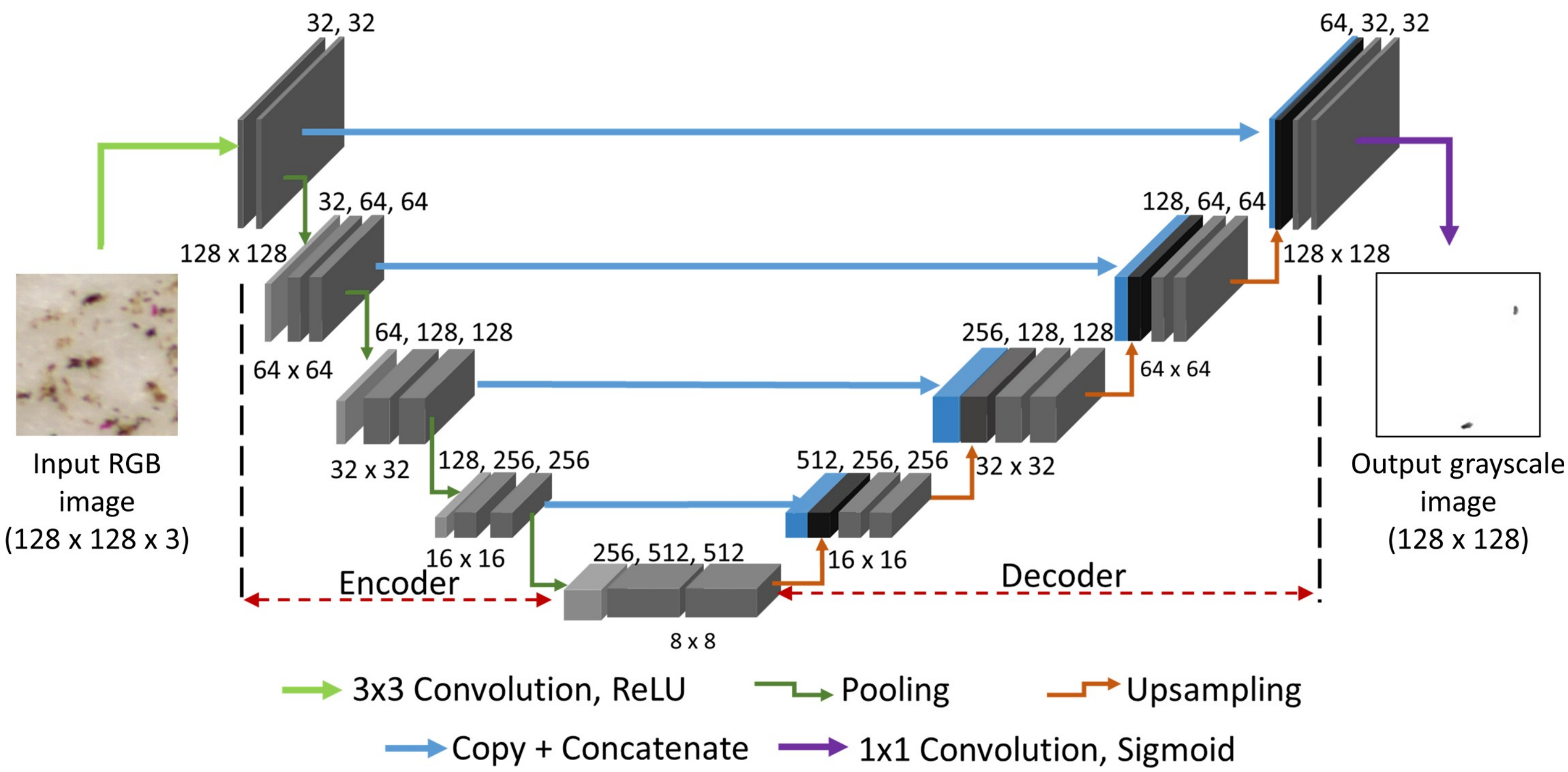

**Fig 2. The convolutional autoencoder network model.** Each box represents a three dimensional feature map (W, H, and D). The depth (D) of each box is mentioned on top and the x-y sizes (W, H) on the bottom. The blue box represents the feature map copied from the encoder step and is concatenated with the feature map (black box) generated by upsampling the previous layer. The input to the network is a 3-channel RGB image and the output is a 1-channel grayscale image. The arrows represent different operations.

## Lensless imaging method

**Hardware and software components.** Fig 1C shows the setup for the lensless egg counting. The basic principle of lensless imaging is discussed elsewhere [20], and has been used to image microscopic objects such as cells, bacteria, and even nematodes without the need for expensive microscopes. Our lensless imaging hardware consisted of the lighting, imaging, and processing modules. (i) The lighting module comprised a light emitting diode (LED, wavelength λ = 616 μm, Vishay Intertechnology) that was aligned with a 100 μm pinhole (Edmund Optics) and sealed within a pinhole mount. This arrangement allowed the light from LED to be emitted through the pinhole producing spatial coherence [20]. The LED was connected in series with a 60 Ω resistor and is powered by the 5 V GPIO (general purpose input/output) pin on the Raspberry Pi (RPi) 3 Model B board. Fig 3A shows the wiring diagram. (ii) The imaging module comprised a Pi camera (8 MP) with a CMOS sensor. The default housing of the camera was removed using a standard razor blade to expose the CMOS sensor. Fig 3A shows the camera connected to the CSI (camera serial interface) port on RPi. (iii) The processing module comprised a portable RPi microcomputer having a 1.2 GHz 64 bit central processing unit (BCM2837), 1 GB random-access memory, and a built-in Wi-Fi unit. The RPi ran on the Raspbian Jessie operating system (Debian "Jessie" based/Linux kernel) loaded onto a 64 GB microSD card (SanDisk™). Camera support was enabled in the settings, and open-source video streaming software was installed to access the live video feed remotely [21]. The RPi was powered by a micro USB wall-mount power supply (5 V, 2.1 A). A 3-D rendering of the platform that housed the various modules is shown in Fig 1C and the platform served to reduce interference from ambient light. The platform was printed on a 3-D printer (da Vinci 1.0 Pro, XYZprinting) using a Polylactic acid filament (1.75 mm diameter, XYZprinting).

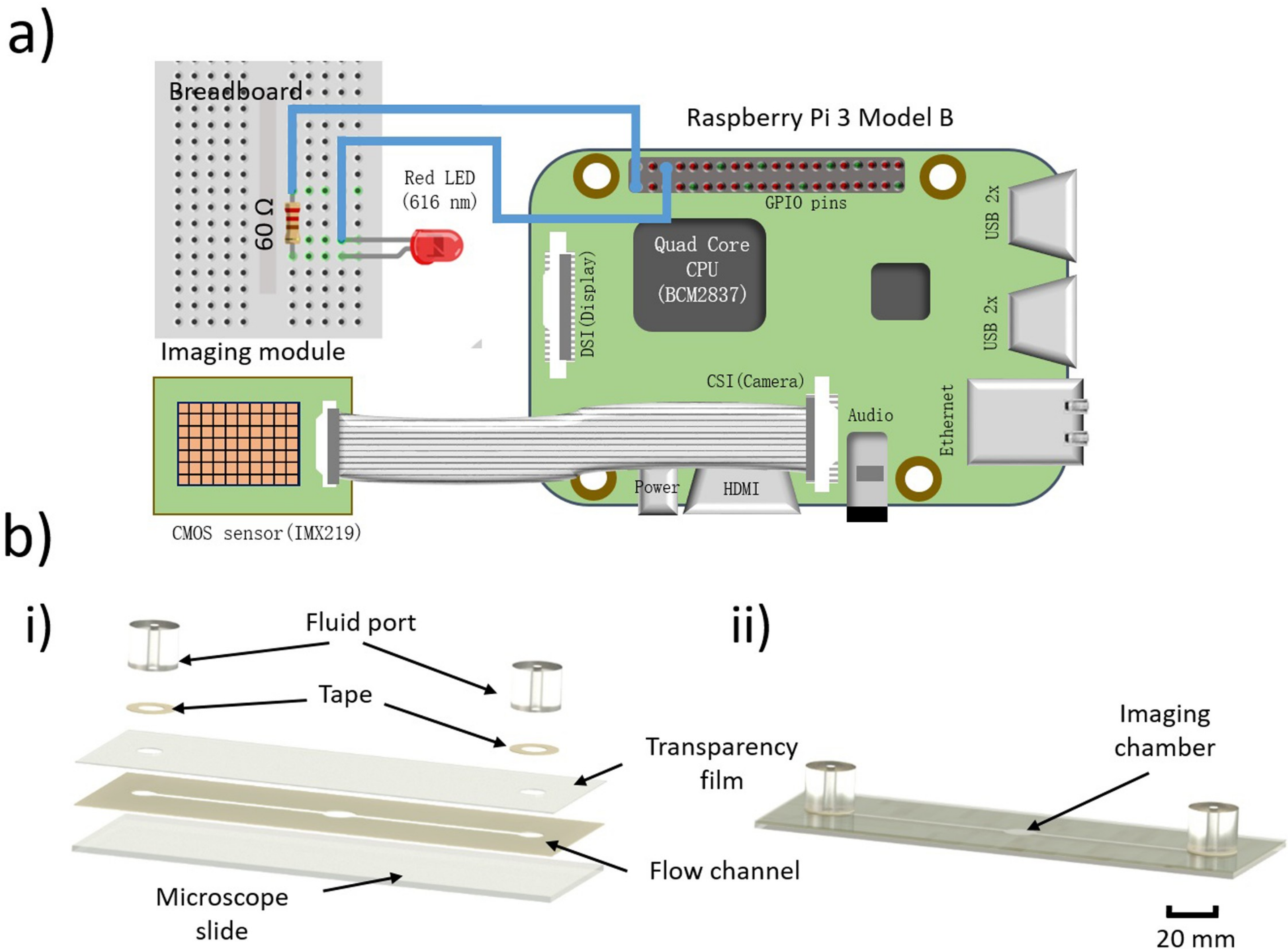

**Fig 3. Wiring diagram for lensless imaging and design of the microfluidic flow chip.** a) The CMOS sensor was connected to RPi through the CSI port. The 5V GPIO pin of RPi was connected to the anode of the LED while the ground pin from RPi was connected to the 60 Ω resistor. The cathode of the LED was connected to the other end of 60 Ω resistor. b) (i) An illustration of the microfluidic flow chip is shown. The microfluidic flow design was cut on a piece of double-sided tape using a cutting machine, and the tape was bonded to a microscope slide. A transparency film with input and output accesses was cut and adhered to the top side of the tape. Two circular tape pieces with holes cut in the center were attached to the fluid ports and placed around the access holes in the transparency sheet. (ii) A diagram of the assembled microfluidic flow chip is shown with all of the layers bonded together. Scale bar = 20 mm.

After sample cleaning as described earlier, 1 mL of the processed sample was loaded in a 3 mL syringe (BD Biosciences) and connected to a syringe pump (KDS-100, KD Scientific). The standard needle of the syringe was replaced by a dispensing needle (16 gauge5.08 cm, Howard Electronic Instruments) to avoid clogging. The syringe was positioned vertically and taped with a vibratory motor (3 V/60 mA, 6500 RPM, Jameco Reliapro) to prevent settling of debris in the tubing. The 3.3 V GPIO pin was used to power the vibratory motor. A plastic tubing (inner diameter: 16 gauge) dispensed the processed sample from the syringe to a microfluidic flow chip. The RPi was turned on and video recording was enabled as the eggs and debris passed through the flow chip. Another piece of plastic tubing was connected to the output port to direct the processed sample into a waste reservoir. After all of the processed sample passed through the flow chip, the video recording was terminated.

**Design of the microfluidic flow chip.** The microfluidic flow chip was intended to be made using low-cost and simple-to-use materials and tools, thereby eliminating the need for micromachining or microfabrication techniques such as photomask design, lithography, spin coating, developing, curing, and etching. The flow chip comprised three layers: the base, channel, and cover layers. A microscope slide (25 × 75 × 1.0 mm, Fisherbrand™) served as the base layer. The channel layer was designed and cut from double-sided tape (25 × 75 × 0.2 mm, 3M™). The cover layer was constructed from a transparency sheet (Apollo™). Initially, commercial software (Studio™, Silhouette America) was used to create the design of the microfluidic flow channel (1.4 × 50 × 0.2 mm) with a central imaging chamber (2.1 × 2.4 × 0.2 mm). The different layers are shown in Fig 3B(i). Double-sided tape was attached to the cutting mat and loaded into the cutting machine (Cameo™, Silhouette America) to cut the flow channel onto the tape. The cut tape was then removed, aligned, and bonded to the base microscope slide. The design of the cover layer had holes for an inlet and an outlet which were aligned with the ends of the flow channel. The same cutting process described above was used to create the cover layer in a transparency sheet, which was placed and bonded to the channel layer. PDMS ports were used as input and output ports and sealed to the transparency film using circular rings of double-sided tape as shown in Fig 3B(i). The assembled flow chip is shown in Fig 3B(ii).

**Data collection and pre-processing.** The videos recorded from the RPi were reconstructed using a custom software program written in Python. As shown in Fig 4, the program loaded one video at a time and read all frames sequentially. The objects of interest (eggs) in a frame were identified by subtracting the current frame from the previous frame and filtering the noise using a median filter. This step eliminated any static content in the current frame including imperfections in the flow channel, settled debris, and image noise. Only the moving objects remained after this step, including eggs and debris. The resultant image was in a binary format with object pixels colored in white against a black background. From the binary image, different physical dimensions (i.e. width, height, shape, and area) of the identified objects were determined and compared with those of a typical SCN egg. This helped identify if the selected object was a SCN egg. The process continued until all of the detected objects in the image were validated. The program then read the next frame and repeated the process until the final frame was reached. Thereafter, the total SCN egg count was reported for the processed sample.

**Principle of reconstructing holographic images.** The spatially incoherent light from the LED passed through an aperture resulting in a partially coherent light called the *reference wave* $U_R(x, y, z)$. The reference wave illuminated the objects located at distance $z_1$ in the channel of the flow chip. The incident light was scattered to generate the *object wave* $U_O(x, y, z)$. The interference between the reference wave and object wave produced holographic patterns. The intensity $I(x, y, z)$ of the holograms at a vertical distance $Z$ is described in Eq 1 [22]:

$$I(x,y,z) = |U_R(x,y,z)|^2 + |U_O(x,y,z)|^2 + U_R^*(x,y,z)U_O(x,y,z) + U_R(x,y,z)U_O^*(x,y,z) \quad (1)$$

The first and second terms in Eq 1 represented the background and scattered light intensity. The scattered intensity is generally weaker than that of the background. The third and fourth terms represented the interference maxima and minima [22]. The CMOS sensor was placed at a distance $z_2$ from the light source ($z_2 > z_1$) and recorded the intensity of the holograms [20,22]. However, the phase information was lost, which made it difficult to discern objects from a digital holographic image.

Digital reconstruction algorithms were used to convert the holographic image to a microscope-like image. The reconstruction commonly known as 'holographic reconstruction' converted the holographic image in the object plane to the detector plane. For our application, we have used the Fresnel diffraction method which utilized a single Fourier transform ($\mathcal{F}$) to back propagate the wave a distance $z = z_2 - z_1$ to the reconstruction plane, $U_z(x, y, z)$. Eq 2 describes how the reconstruction plane was obtained by Fourier transforming the element-wise multiplication of image intensity and combining it with a Fresnel approximated transfer function and a phase term. This operation can be written as [23]:

$$U_z(x, y, z) = \frac{e^{jkz}}{j\lambda z} e^{\frac{j\pi}{\lambda z}(x^2+y^2)} * \mathcal{F}\left\{ I(x, y, z) * e^{\frac{j\pi}{\lambda z}(x^2+y^2)} \right\} \tag{2}$$

where $\lambda$ is the wavelength and $k = \frac{2\pi}{\lambda}$ is the wave number. The pixel size $\Delta\eta$ in the reconstruction plane was directly proportional to the reconstruction plane as shown in Eq 3.

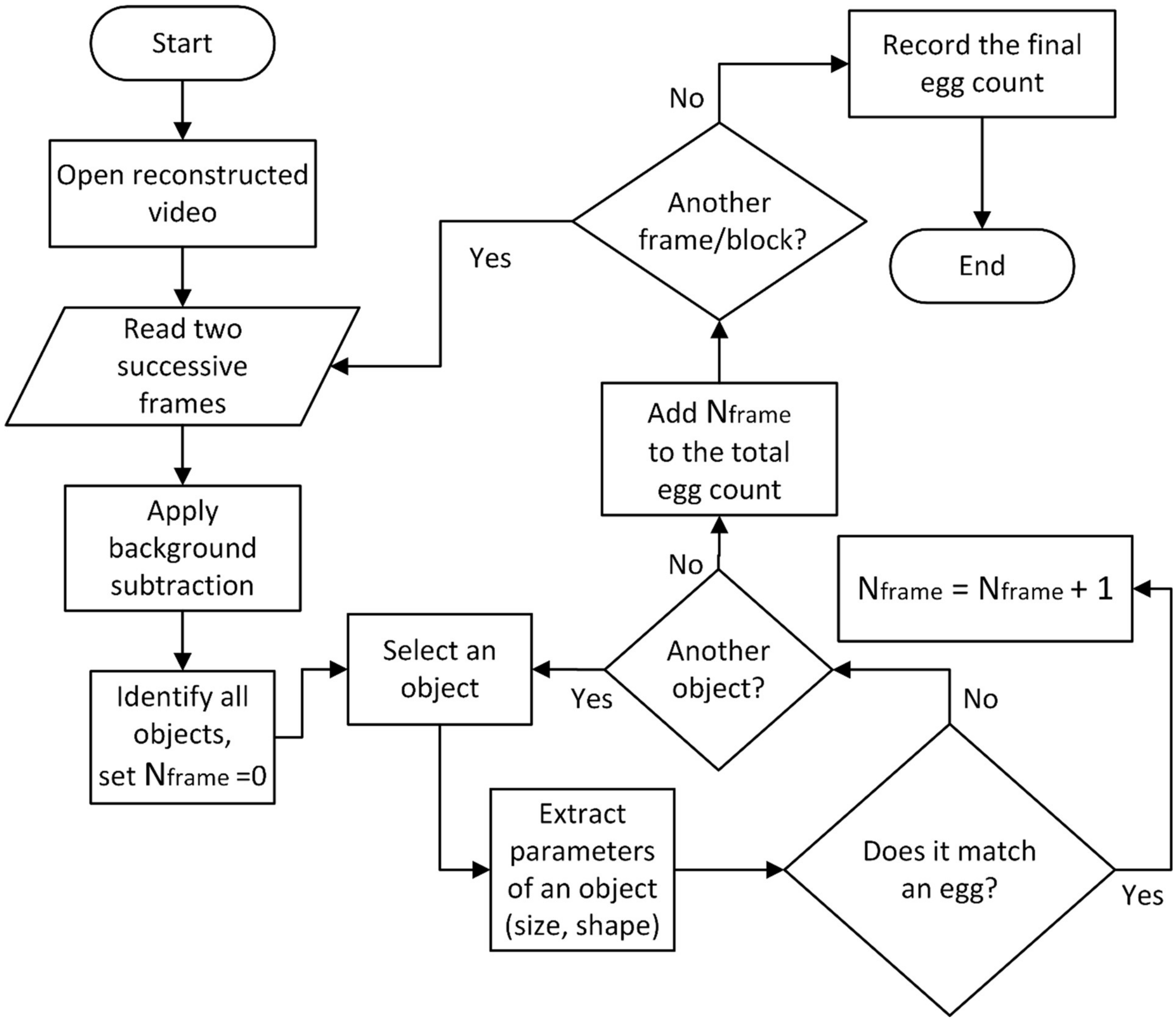

**Fig 4. The algorithm for the lensless SCN egg counting method is depicted as a flow chart.**

$$\Delta\eta = \frac{\lambda z}{N\Delta_p} \tag{3}$$

where $\Delta_p$ is the pixel pitch of the CMOS sensor and N is the minimum dimension of the holographic image size. The reconstructed images at this stage often looked blurred because of the presence of zeroth order frequencies, which can be removed if the phase is known. An approximate phase was obtained using the Fresnel approximation. There are other techniques to recover the phase that involve taking multiple images of the same object by varying parameters such as wavelengths, angles of the LED, and z values [24,25]. Some other techniques use iterative methods to retrieve the phase. These alternatives are also effective but require additional imaging components and computations.

## Results

### OptiPrep™-based density gradient centrifugation

A qualitative analysis of the debris distribution was performed to characterize the effects of OptiPrep™ during the centrifugation step. Using the protocol described in Fig 1A, different OptiPrep™ solutions (0%, 20%, 40%, 50%, 60%, and 80% by volume) were used. Images of the centrifuge tubes before and after the centrifugation step were taken for the different OptiPrep™ solutions and cropped as shown in Fig 5 (i,ii). The three layers were separated after centrifugation, diluted with water to 12 mL, and put on separate filter papers. Fig 5 (iii) shows a plot of the mean pixel intensity levels as a function of distance along the center axis of the tube. In these plots, the orange and purple lines denote the mean pixel intensity levels before and after the centrifugation step, respectively.

Compared to the control treatment (i.e. 0% OptiPrep™) (Fig 5 Control), the sample cleaning step using OptiPrep™ resulted in a marked visual separation of the debris. Before centrifugation, the tube had a relatively consistent debris distribution (Fig 5 Control i.) whereas after centrifugation, all of the suspended debris were concentrated into the pellet at the bottom of the tube (Fig 5 Control ii.). Since the control sample did not contain the OptiPrep™ solution, there was no distinct interface layer. Here the mean intensity plots before and after centrifugation are similar because no OptiPrep™ solution was present (Fig 5 Control iii.).

With 20% OptiPrep™, there was a clear contrast between the images of the tube before (Fig 5 20% i.) and after centrifugation (Fig 5 20% ii.). The mean pixel intensity plot showed a definitive change at the interface before and after centrifugation (Fig 5 20% iii.). The interface layer for 40% and 50% OptiPrep™ appeared visually similar to that of the 20% concentration. However, with increasing OptiPrep™ concentrations (60% and 80%), there was a noticeably increasing amount of debris trapped at the interface layer both before and after centrifugation (Fig 5 60% i.-ii. and Fig 5 80% i.-ii.). This separation of debris at the interface layer was consistent with the mean color intensity plot (Fig 5 80% iii.).

### Scanner-based method: Egg recovery in the interface layer

Each image of the filter paper corresponding to a different OptiPrep™ solutions (20%, 40%, 50%, 60%, and 80%) was run through a custom software program, written in Python, to identify the nematode eggs and give a count of the eggs within the image. The program initially takes a full image of the filter paper and then partitions it into patch images and sub-patches to identify and count the eggs. Fig 6A and 6B show representative images of two filter papers having an uncleaned sample and a cleaned sample, respectively. In both cases, the program was able to differentiate stained eggs from the debris.

For a given volume of the processed sample, the total SCN egg count was obtained by combining the egg counts for the top, interface, and bottom layers. The egg recovery ratio was calculated as a ratio of the egg count in each of the three layers to the total egg count as depicted in Fig 7. This was repeated for four processed samples (n = 4). The average egg recovery ratio and standard deviation were plotted for the top, interface, and bottom layers corresponding to the different OptiPrep™ solutions.

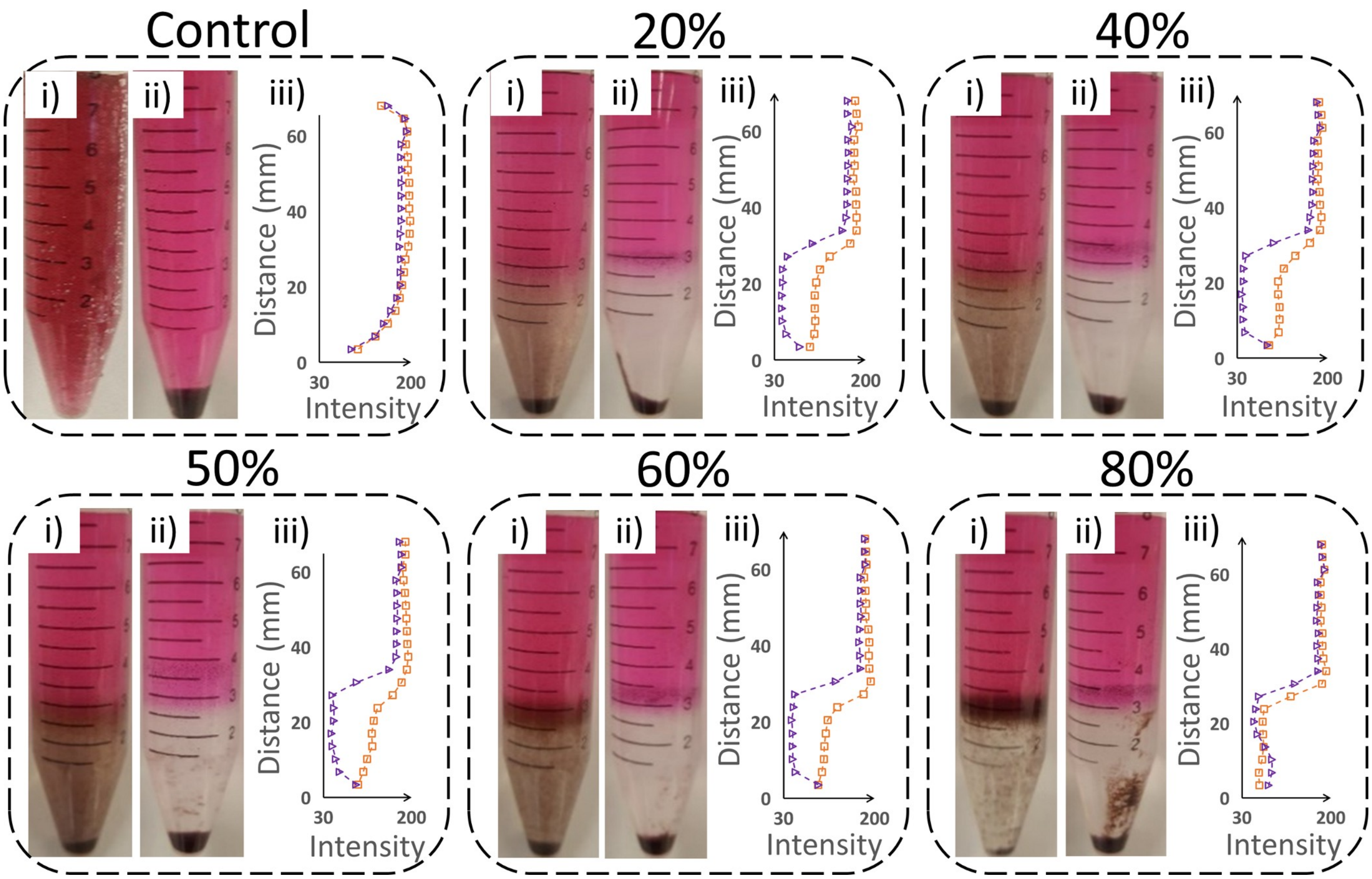

**Fig 5. Distribution of debris in the OptiPrep™ solution before and after centrifugation.** (i., ii.) Cropped images of the centrifuge tube before and after centrifugation. iii.) The mean pixel intensity of the samples in the centrifuge tube as a function of vertical distance (in mm) was plotted for before (orange) and after (purple) centrifugation. The different OptiPrep™ solutions used were 0%, 20%, 40%, 50%, 60%, and 80% by volume. The control (i.e. 0% OptiPrep™) does not have an interface layer, and so the supernatant is treated as the top and the pellet as the bottom layer. All test tubes pictured were 15 mL tubes.

https://doi.org/10.1371/journal.pone.0223386.g005

The graph in Fig 7A shows that the egg recovery ratio in the top layer was negligible after the OptiPrep™ sample cleaning as virtually no eggs were found in this layer. In the interface layer, the egg recovery ratio improved with increasing OptiPrep™ concentration and approached 95% for the case of the 80% concentration of OptiPrep™. Meanwhile, the egg recovery in the bottom layer decreased with increasing OptiPrep™ concentrations and approached 4% for the case of the 80% OptiPrep™ solution. For the remaining studies with OptiPrep™, the 50% OptiPrep™ solution was used as the egg recovery at the interface layer was greater than 80% and the sample purity was the highest. Beyond an OptiPrep™ concentration of 50%, sample purity started to decrease as was demonstrated by the increase in material remaining at the interface layer after cleaning (Fig 5 60% ii.).

The graph in Fig 7A shows that most eggs were recovered from the interface layer of the centrifuge tube after sample cleaning with OptiPrep™. Fig 7B tabulates the average percentage of egg recovery at the interface layer for the associated volume percent OptiPrep™ solutions (n = 4). The spatial distribution of eggs in the images of the filter papers (corresponding to the interface layer) are shown as raster plots in Fig 7C. Four extraction runs were conducted, each with a fresh batch of extracted and stained SCN eggs and a pre-specified OptiPrep™ solution. The raster plots show the distribution of the SCN eggs (denoted as dark pixels) across the 324 sections of the images generated by the software. The raster plots show that the eggs (or dark pixels) were distributed across most of the filter paper, which helped facilitate the identification of individual eggs. Comparing the raster plots for 20% OptiPrep™ with those of higher OptiPrep™ concentrations, the raster plots appear darker as the OptiPrep™ concentration increases, suggesting that the egg recovery ratio increased in the interface layer as OptiPrep™ concentrations increased.

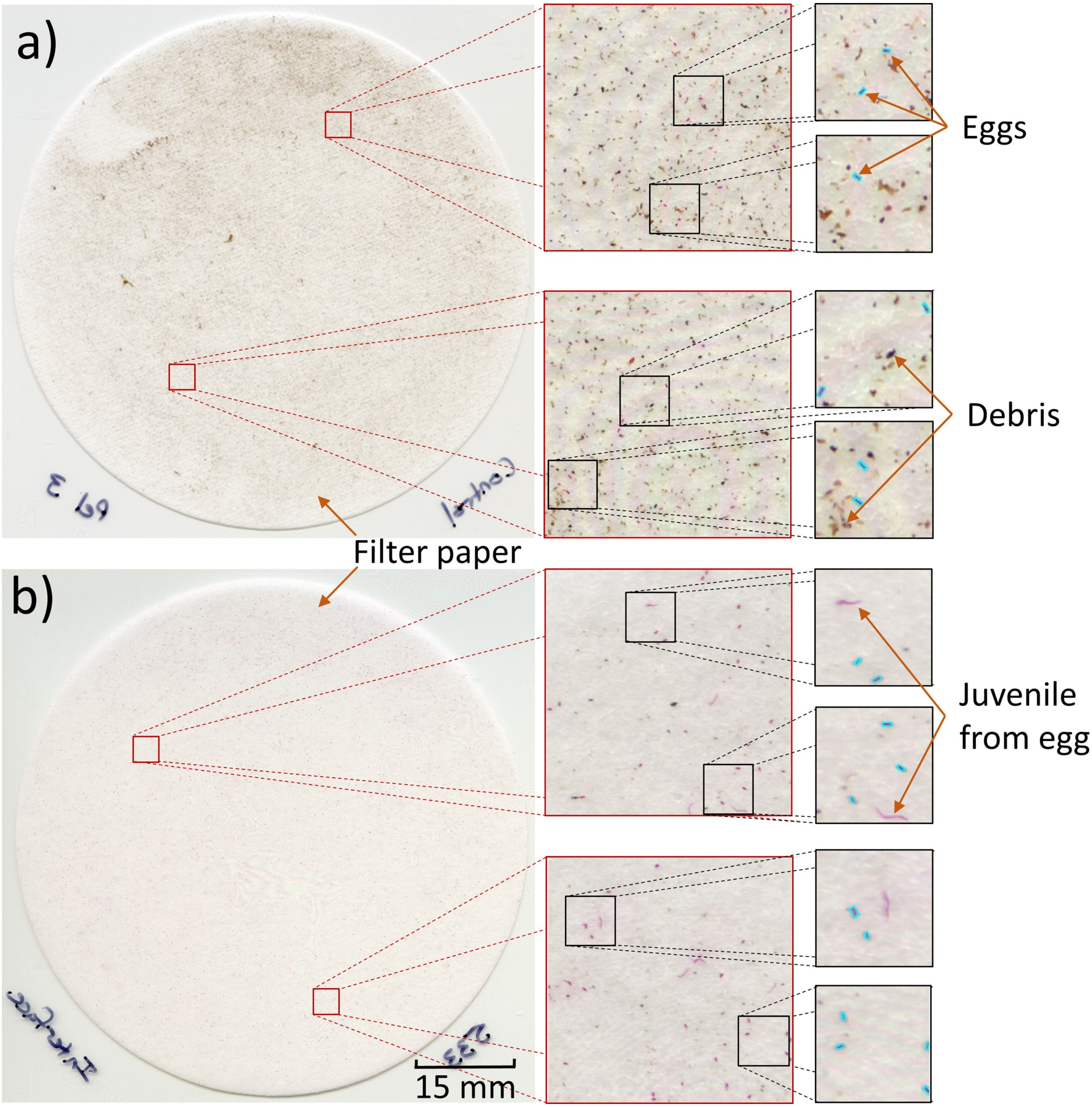

**Fig 6. Software program to automatically detect SCN eggs among the debris in the scanner-based method.** The software program starts with the full image of the filter paper and partitions it into patch images and sub-patches to identify and count the SCN eggs. a) An example of identifying eggs in a debris-laden, uncleaned sample is shown. b) An example of identifying eggs in an OptiPrep™ cleaned sample is shown where it is even possible to differentiate between stained eggs and juvenile nematodes which have recently emerged from the eggs. Scale bar = 15 mm.

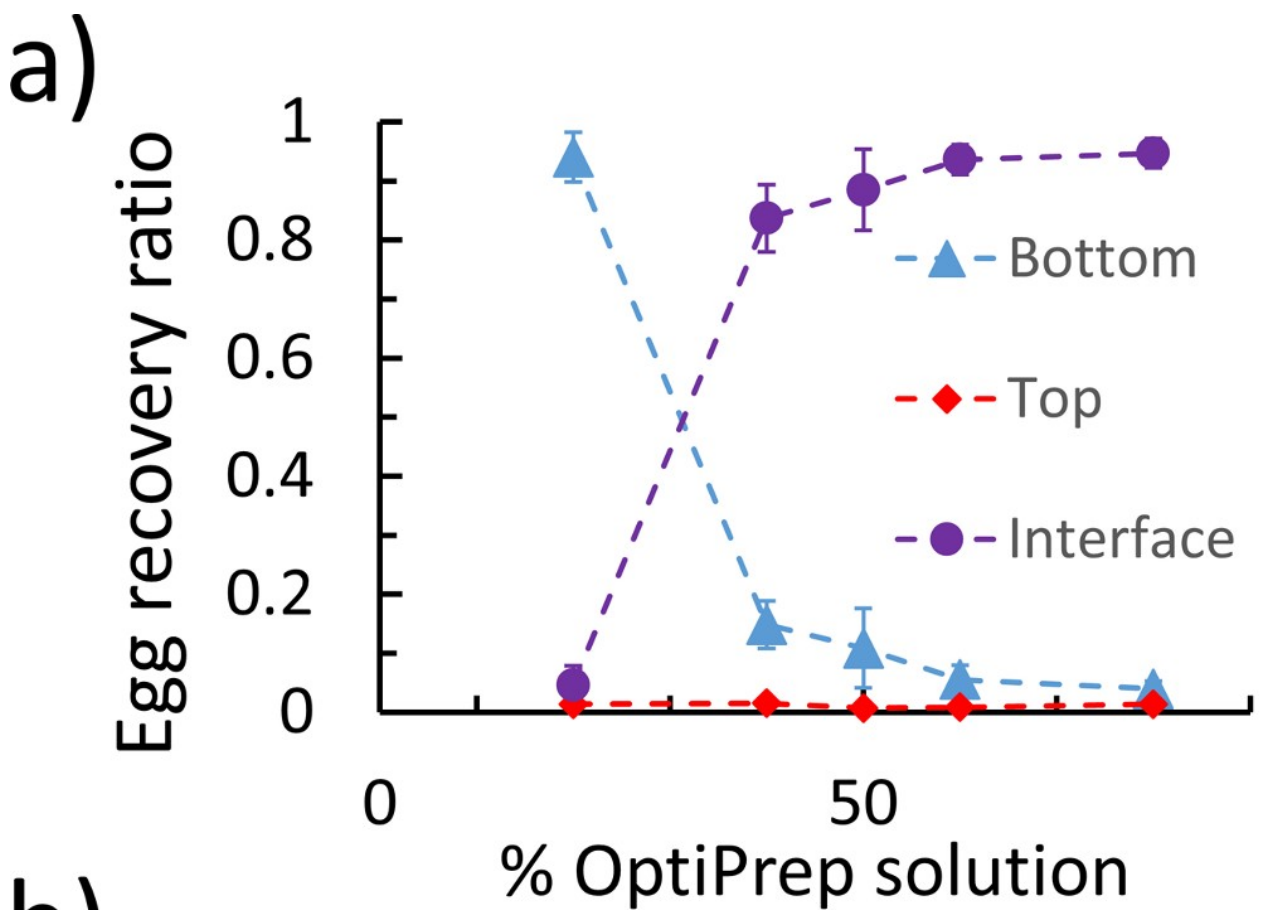

b)

| OptiPrep volume percentage | Interface egg recovery (%) |
|---|---|
| 20 | 4.6 ± 3.2 |
| 40 | 83.7 ± 5.7 |
| 50 | 88.5 ± 6.8 |
| 60 | 93.6 ± 2.5 |
| 80 | 94.7 ± 2.5 |

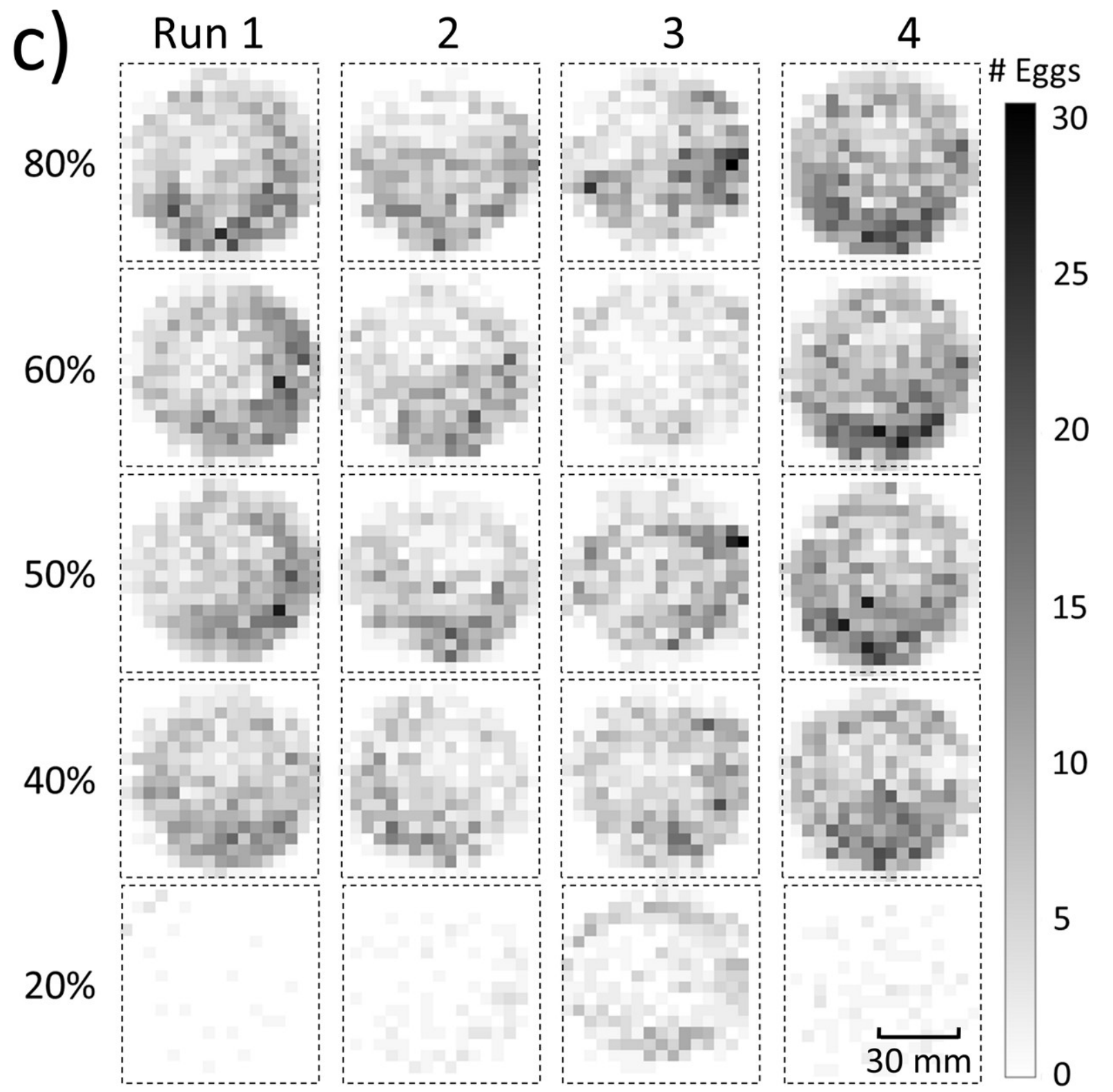

**Fig 7. Sample cleaning using OptiPrep™ and SCN egg recovery from the three layers.** a) The graph shows the ratio of SCN eggs recovered from the three layers (top, interface, and bottom layers) at different concentrations (volume percentages) of OptiPrep™. At 50% and higher OptiPrep™ concentrations, the egg recovery ratio was greater than 80% in the interface layer. b) The egg recovery percentage in the interface layer is listed for different percentages of OptiPrep™ by volume for four separate runs of processed samples. c) Raster plots help to visualize the spatial distribution of the eggs on the filter papers (corresponding to the interface layer) across four extraction runs as the volume percentage of OptiPrep™ was varied. The egg count for each patch image is denoted as a dark pixel in the raster plot.

## Performance of sample cleaning methods: OptiPrep™ versus sucrose

The OptiPrep™ sample cleaning method was compared with the sucrose centrifugation technique, which is the conventional method of separating nematodes and eggs from debris [26]. The samples were treated as described in Fig 8(A). A stock sucrose solution was made by dissolving 454 gm of sucrose in 1 L of tap water. Approximately 500 $cm^3$ of SCN infested soil was manually processed (using techniques in the 'Sample Preparation' section) to extract the SCN cysts, then eggs and debris. The samples (i.e. liquid containing eggs and debris) were diluted to 500 mL and mixed thoroughly before obtaining the manual egg counts for this bulk sample using a nematode counting slide and dissecting microscope with a magnification of 100×.

The bulk egg sample was stained with acid fuchsin and split into ten different 50 mL test tubes. The contents in five of these test tubes were cleaned using the OptiPrep™ cleaning method (using a 50% by volume OptiPrep™ solution). The egg counts for these five samples were obtained using the lensless method and the scanner-based egg counting method. The egg counts were reported as the number of eggs per milliliter.

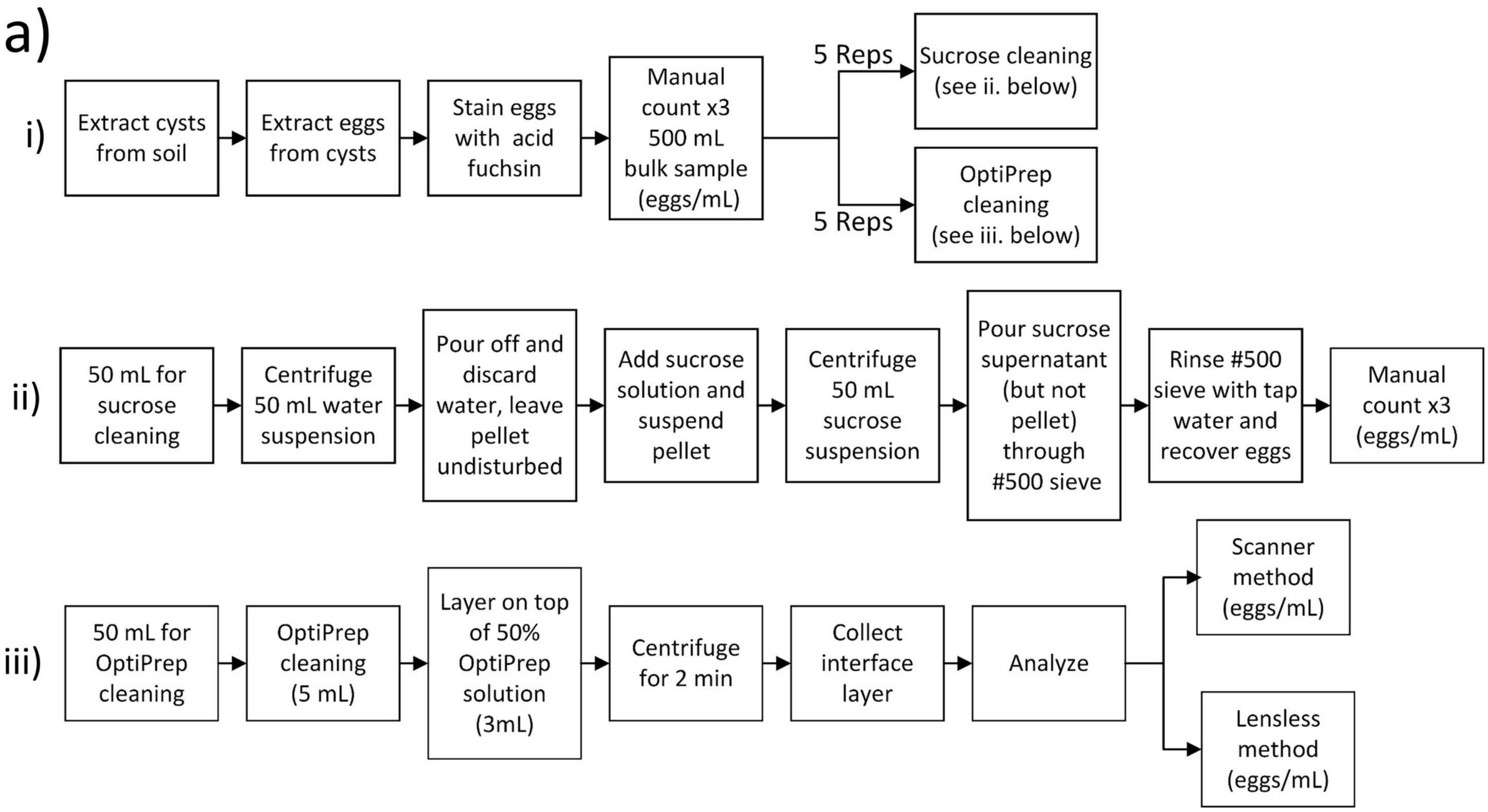

b)

| Farm Location | Bulk[a] (Nematode counting slide) | Sucrose[a] (Nematode counting slide) | OptiPrep[a] (Scanner) | OptiPrep[a] (Lensless) |
|---|---|---|---|---|
| Muscatine County, IA | 129 ± 21 | 55 ± 4 | 99 ± 3 | 79 ± 17 |
| Story County, IA | 41 ± 9 | 20 ± 2 | 44 ± 4 | 40 ± 7 |

[a]Counts in eggs/mL

**Fig 8. Comparison of the sample cleaning and SCN egg counting methods.** a) i-iii) Steps performed to extract cysts, then eggs from cysts, stain eggs, perform sample cleaning (OptiPrep™ and sucrose cleaning methods), and egg counting (manual, scanner-based, and lensless imaging). b) The table lists the average egg count (n = 5) and standard deviation for the egg count from scanner-based and lensless methods compared to the standard nematode slide count on the uncleaned bulk sample.

The remaining five test tubes of samples were cleaned by the sucrose centrifugation technique. Each tube was centrifuged for 5 minutes at 420 G and the excess water was decanted off. The pellet was thoroughly mixed in 50 mL of the stock sucrose solution and centrifuged for one additional minute to separate the eggs from the soil pellet. The sucrose solution supernatant, containing the eggs, was poured over a 37-μm-pore sieve and rinsed in water to remove the excess sucrose. Three 1 mL sub-samples of the processed egg samples were placed on a nematode counting slide, and the SCN eggs were manually counted under a microscope and averaged to estimate the egg count per milliliter.

Soil samples were collected from two fields in Iowa–one each in Muscatine and Story Counties. Soil was collected from two different places in Iowa to get a range of soil textures with which to test the egg cleaning methods. Both methods of sample cleaning (i.e. OptiPrep™ and sucrose based) were conducted on ten different samples (five from Muscatine and five from Story). The egg counting was done by three methods–nematode counting slide, scanner-based, and lensless based. The mean and standard deviation of the egg count obtained from the different sample cleaning methods (i.e. bulk, sucrose, OptiPrep™) and counting methods (i.e. nematode counting slide (Chalex, LLC, Park City, UT), scanner, lensless setups) are tabulated in Fig 8(B).

## Lensless imaging method: Reconstruction of holographic videos

After sample cleaning using 50% OptiPrep™ by volume, a 1 mL solution of the interface layer was loaded into a syringe. The dispensing needle of the syringe was connected to the input port of the microfluidic flow chip. The syringe pump was turned on at a flow rate of 1 mL per hour. As the liquid and particles began to pass through the flow channel, the RPi was turned on and the holographic video was recorded (S1 Video). At the end of the video recording, the video was transferred to a remote workstation and processed by a MATLAB script to generate the reconstructed video. Reconstruction of a single frame of the video took approximately 4 seconds on the CPU (Intel Xeon E5, 32GB RAM).

Fig 9A illustrates the enhancement in object clarity after image reconstruction. Fig 9A(i.) shows a section of the image captured from the lensless imaging setup (i.e. raw image). After reconstruction, the clarity and focus of the raw image was considerably enhanced, and the reconstructed image is shown in Fig 9A(ii.). For comparison, the bright-field image of the same area was captured by a stereo microscope at 50× magnification and is shown in Fig 9A(iii.).

The reconstructed video had sufficient resolution and clarity to identify the nematode eggs in every frame. Fig 9B shows time-lapsed image frames of a representative video that was recorded from the lensless imaging setup and reconstructed thereafter. The image frames refer to a small area of the flow channel. The eggs detected in each frame are marked in cyan color with their distinct egg number. The video was recorded at 1 frame per second and the flow rate was set to 1 mL/hour (S2 Video).

## Performance of automated egg counting methods

We compared the accuracy of the egg counts obtained from the software program with manual egg counts as shown in Fig 10. In total, there were fifteen images of filter papers corresponding to the five different OptiPrep™ concentrations and their three individual layers (i.e. top, interface, and bottom). Initially, the images of filter papers were divided into 324 patch images (1024 × 1024 pixels) and saved in separate folders on the computer. To obtain the manual egg counts, a user viewed each patch image and counted the number of eggs (Fig 8). This process was repeated for all fifteen of the filter paper images. Next, to obtain egg counts from the software program, each patch image was further split into 256 × 256 pixel sub-patches and analyzed using the trained deep learning model. Since the deep learning model was trained for 128 × 128 pixel images, the sub-patches were resized before being passed to the learning model. Fig 10A shows the total egg counts calculated by the software program and by the manual method for all the filter paper images. The $R^2$ value between these two sets of data was high indicating that there was good agreement between the two sets of counts. We also investigated the correlation between the software and manual egg counts at a specific OptiPrep™ concentration (50%). There were a total of 972 patch images across the three layers, and the correlation between them was high as well, as shown in Fig 10B. This high correlation demonstrates that our software program can provide the egg count in processed samples with accuracy that is comparable to visual detection.

To test the correlation of the software for egg counting used with the lensless imaging setup, multiple samples (n = 12) were used. The processed samples, after cleaning with 50% OptiPrep™, were passed through the lensless imaging setup. The raw holographic videos were

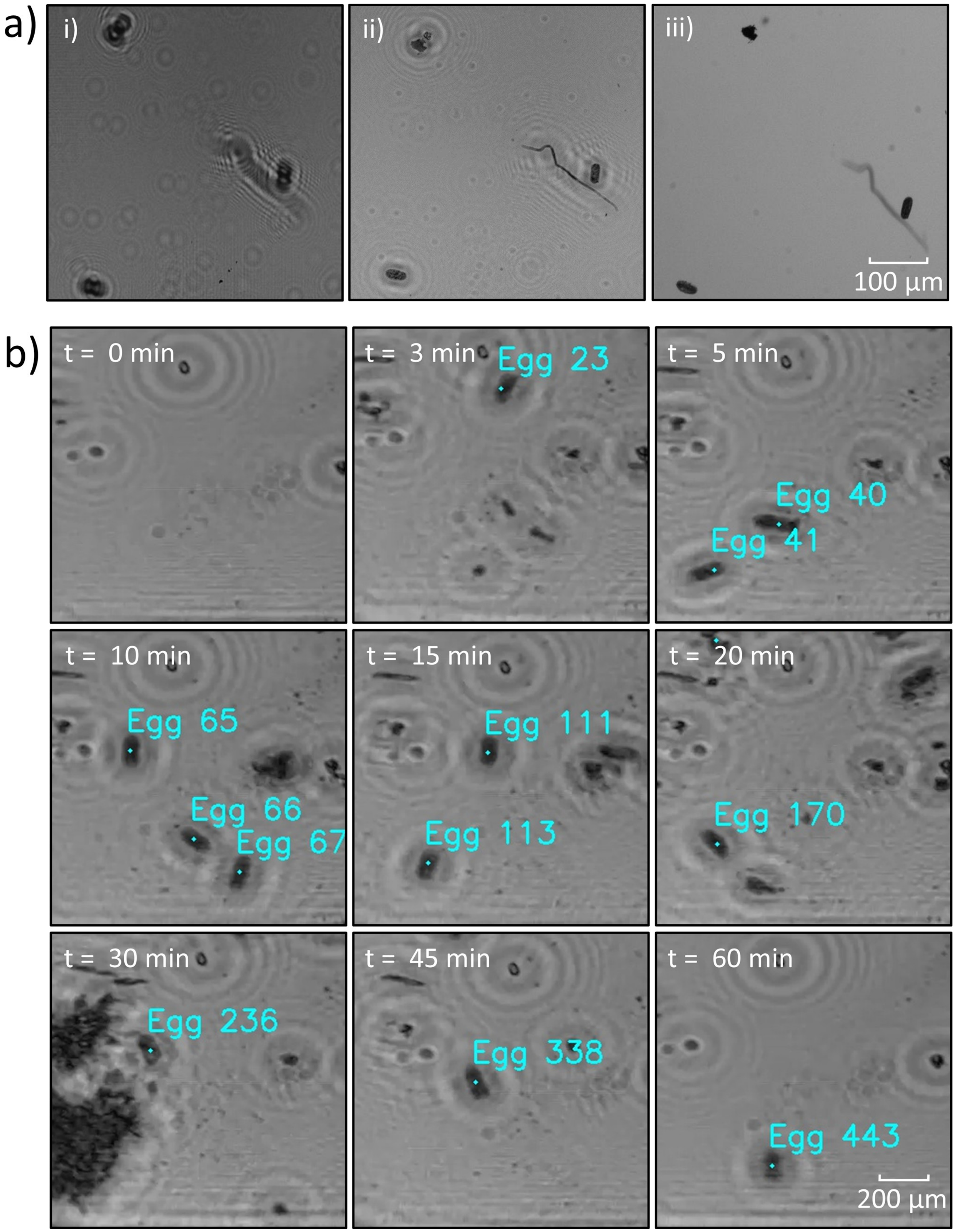
a)
i)
ii)
iii)
100 µm
b)
t = 0 min
t = 3 min
Egg 23
t = 5 min
Egg 40
Egg 41
t = 10 min
Egg 65
Egg 66
Egg 67
t = 15 min
Egg 111
Egg 113
t = 20 min
Egg 170
t = 30 min
Egg 236
t = 45 min
Egg 338
t = 60 min
Egg 443
200 µm

**Fig 9. Reconstruction of the holographic videos and a time-lapsed visualization of the lensless imaging method.** a) i) The raw image of a sample area was recorded by the lensless imaging setup (z = 8 mm). ii) The reconstructed image was produced using the Fresnel diffraction method. iii) The sample area was imaged with a stereo microscope with bright-field illumination at 50× magnification. Scale bar = 100 μm. b) The images were taken from a representative holographic video recorded by the lensless imaging setup and reconstructed thereafter. A small area of the flow channel is shown here. The eggs detected by the software were labeled with a distinct egg number. Scale bar = 200 μm.

recorded as described earlier. After reconstruction, the videos were run through the software program to identify the eggs in every frame and provide the total egg count for the entire video. The processed sample was retrieved at the outlet of the flow channel and passed through the scanner-based egg counting method. The egg counts from the lensless method and the scanner-based method were plotted for all the samples as shown in Fig 10c. There was high correlation between the egg counts from both the methods with egg counts determined by direct microscopic observation.

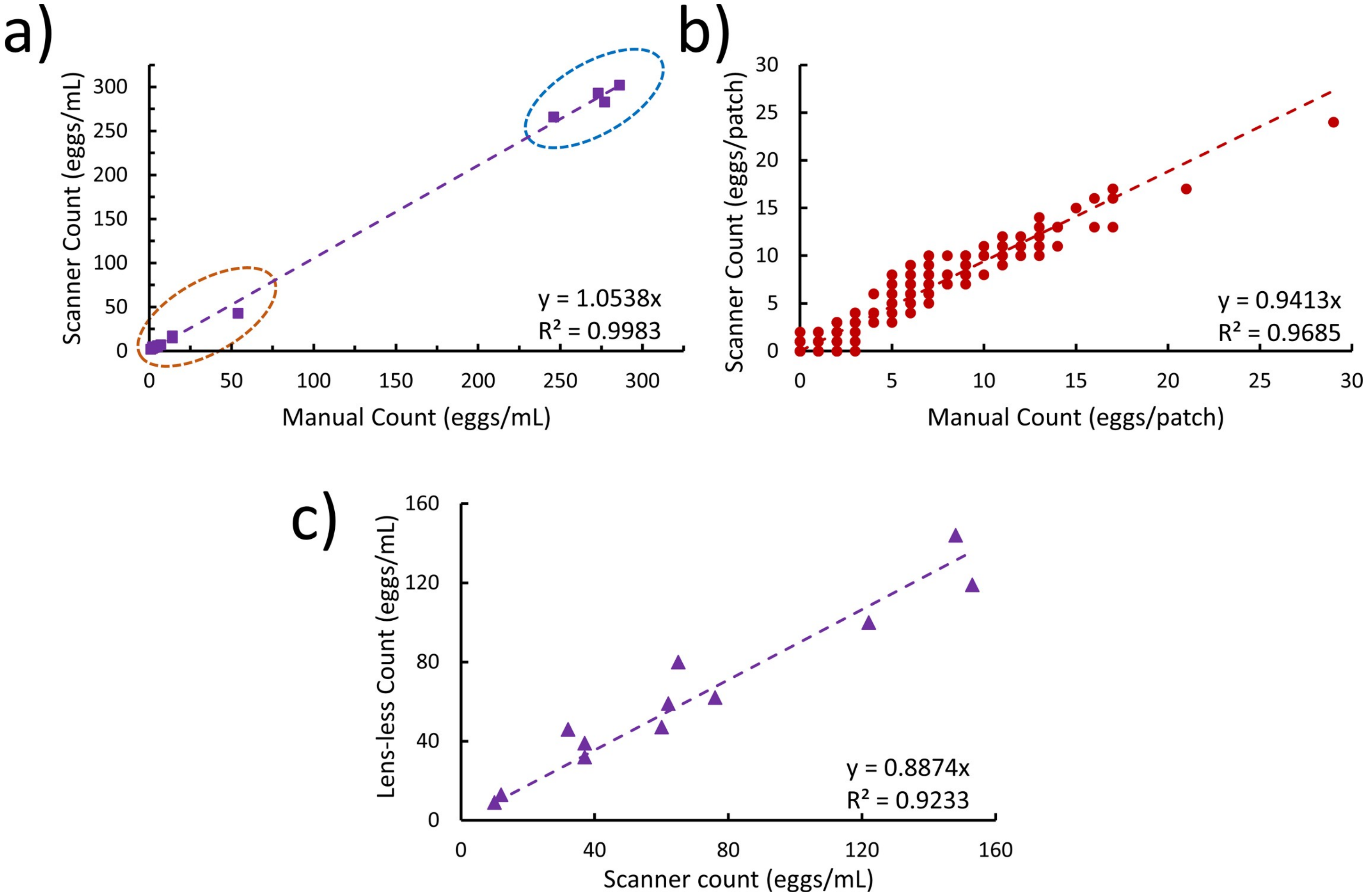

**Fig 10. Performance of the software programs for automated SCN egg counting.** a) The plot shows the SCN egg counts obtained for the three layers (top, interface, and bottom) for all different OptiPrep™ concentrations. The data points circled in orange belong to samples with fewer eggs. Similarly, the points circled in blue belong to samples with a large number of eggs. b) The plot shows the correlation between manual and software egg counts for all image patches of the three layers at 50% OptiPrep™ by volume. The data were collected from three filter papers, each subdivided into 324 patch images. c) The egg counts obtained from the two methods are compared: software program for lensless imaging method and the scanner-based egg counting method. Each processed sample was passed through the lensless imaging setup and further analyzed by the scanner-based method. There is high correlation between the egg counts from the two methods. The plotted data corresponds to 12 different processed samples.

## Discussion

### OptiPrep™-based sample cleaning method

The current state-of-the-art sample cleaning method involves the use of sucrose centrifugation to separate plant-parasitic nematode juveniles and eggs from debris. While sucrose is inexpensive and readily available, its high osmotic potential may have adverse effects on nematode juveniles and eggs upon prolonged exposure. Also, if the sucrose solution is not adequately washed away from the nematodes and eggs, there is the risk of bacterial and fungal growth in the samples. To address these challenges of sucrose centrifugation, we developed an alternative density gradient method using OptiPrep™ solution to separate plant-parasitic nematode eggs from debris. OptiPrep™ has certain advantages over sucrose. OptiPrep™ is iso-osmotic and nontoxic to biological materials such as cells and tissues. OptiPrep™ has a lower viscosity compared to sucrose, which eases the mixing and washing steps with water. The lower viscosity and osmolality of OptiPrep™ is advantageous to clean small volumes of sample homogenates (5–10 mL in a 15 mL test tube) and to condense the eggs into the interface layer (2 mL) within the centrifuge tube. Greater nematode egg recovery is achieved from 40% to 80% OptiPrep™ solutions (i.e. $>$ 80% recovery) compared to those from the sucrose centrifugation (i.e. less than 40% recovery) (Fig 7). As a disadvantage, the price of the OptiPrep™ density gradient medium is higher (approximately $250 USD for 250 mL) than sucrose. The high price of OptiPrep™ may be a challenge for soil processing facilities which perform large-scale or routine separation of nematode eggs from debris.

### Automated nematode egg counting methods

The traditional method of nematode egg counting is based on microscopic observation of samples. The chamber of nematode counting slides is loaded with 1 mL of the sample, and the number of eggs on the grid is counted manually. Counting nematode eggs in this manner is a cheap and relatively simple procedure but requires considerable time, expertise, and patience of a trained expert [26]. Also, cross contamination can occur because of insufficient or improper cleaning of the chamber in the nematode counting slide between samples. A deep learning architecture was recently developed to identify SCN eggs from microscopic images of nematode-counting slides [27]. Their input images (480 × 640 pixels) were subdivided into 16 × 16 image patches and passed through a selectivity function to identify patches with an SCN egg. However, with this deep learning architecture, it is challenging to handle situations with larger image sizes or multiple eggs (i.e. greater than two eggs) within an image patch such as those obtained from bulk samples on our high-resolution scanner.

The new, scanner-based egg counting method we developed has advantages over previous counting methods. Here the filter papers are automatically scanned to obtain the egg counts using a trained deep learning model. Our deep learning model handles large input images (17759 × 17759 pixels) which are subdivided into 4900 patches (256 × 256 pixels). The U-Net convolutional autoencoder model can recognize multiple eggs within a patch using object localization. The time to obtain the computerized egg counts from a single filter paper image is around 5 minutes, whereas this time is much longer for manual counting and varies across different samples and persons. While the size of the sample assessed using a nematode counting slide is limited by the volume of the observation chamber (1 mL) of the slide, each filter paper can process around 12 mL of the sample suspension. Also, cross contamination between samples is eliminated in the scanner-based method as the substrate is a disposable filter paper. The filter paper substrate is relatively inexpensive (approximately $0.48 USD) compared to the nematode counting slide (approximately $50 USD). As a disadvantage, the scanner-based method requires a pre-processing step where the sample needs to be stained for proper identification. The staining process is performed at high temperature which eventually kill the eggs. Also, it is difficult to recover the eggs after imaging as the processed sample is distributed on a filter paper. Hence the scanner-based method is not suited for experiments where samples containing live eggs or nematodes are to be counted for further use, such as to infest soil for greenhouse experiments.

The new, lensless egg counting method has advantages over the manual slide counting and scanner-based methods. Here egg staining is not required as the lensless imaging setup can enumerate both stained and unstained samples. Because of continuous fluid flow through the ports, the counted eggs are easily recovered for later use from the microfluidic flow chip. The material cost is relatively low (less than $100 USD) for the lensless imaging setup and adhesive tape microfluidics compared to the costs of standard microscopes, scanners, and nematode counting slides used in previous methods. The lensless imaging method is functionally more automated than the scanner-based method as a user only needs to load a syringe containing the egg sample into a syringe pump, set the right volume and speed parameters, and start the video recording process. As a disadvantage, the lensless egg counting method has a relatively long processing time. The flow rate of the sample is 1 mL per hour; higher flow rates produce noise in the recorded images. The reconstruction of the holographic video takes a few hours and the egg counting software needs another 10 minutes to obtain the egg count.

## Conclusion

In conclusion, we developed methods to improve the efficiency of SCN egg counting by sample cleaning and imaging on scanner and lensless setups. We also created software programs to count the eggs. The use of OptiPrep™ during density gradient centrifugation helped capture most of the nematode eggs in the interface layer. Egg recovery was greater than 80% in the interface layer for the case of 50% OptiPrep™ by volume. Then the processed samples (i.e. eggs with debris) were either placed on filter papers to record static images with a flatbed scanner or passed through a microfluidic flow chip to record real-time, holographic videos by a lensless imaging setup. In the scanner-based method, a convolutional autoencoder network recognized the nematode eggs from static scanner images with reasonable accuracy as confirmed by visual observations. For the lensless method, the software program successfully reconstructed the holographic videos and identified the nematode eggs in the processed sample, which was subsequently confirmed by the scanner-based method. The cost of the materials was low by using off-the-shelf components (such as CMOS sensor, Raspberry Pi, LED), microfluidic flow chips made from inexpensive double-sided adhesive tapes, and standard filter papers. Furthermore, the use of image processing and deep learning tools circumvents the need for hiring and training personnel to count nematode eggs using nematode counting slides. With our new methods, manual intervention is only needed during sample preparation and loading; the software programs handle the remaining operations automatically (i.e. image or video capture, detection and counting, data storage) over a broad range of nematode egg numbers (10–300 eggs/mL).

## Supporting information

**S1 Appendix. Details of deep learning model.**
(DOCX)

**S1 Video. Raw holographic video.**
(MP4)

**S2 Video. Reconstructed holographic video.**
(MP4)

## Acknowledgments

This work is partially supported by the U.S. National Science Foundation [NSF IDBR-1556370] to S. P. and G. T. and the Defense Threat Reduction Agency [HDTRA1-15-1-0053] to S. P. We thank David Soh (ISU Department of Plant Pathology and Microbiology) for providing the manual counts of the soil samples.

## Author Contributions

**Funding acquisition:** Gregory Tylka, Santosh Pandey.

**Writing – original draft:** Upender Kalwa, Christopher Legner, Elizabeth Wlezien, Gregory Tylka, Santosh Pandey.

## References

1. Allen TW, Bradley CA, Sisson AJ, Byamukama E, Chilvers MI, Coker CM, et al. Soybean yield loss estimates due to diseases in the United States and Ontario, Canada, from 2010 to 2014. Plant Heal Prog. 2017; 18: 19–27. https://doi.org/10.1094/PHP-RS-16-0066
2. Gerdemann JW. Relation of a large soil-borne spore to phycomycetous mycorrhizal infections. Mycologia. 1955; 47: 619. https://doi.org/10.2307/3755574
3. Byrd DW, Barker KR, Ferris H, Nusbaum CJ, Griffin WE, Small RH, et al. Two semi-automatic elutriators for extracting nematodes and certain fungi from soil. J Nematol. 1976; 8: 206–12. Available: http://www.ncbi.nlm.nih.gov/pubmed/19308224%0Ahttp://www.pubmedcentral.nih.gov/articlerender.fcgi?artid=PMC2620186 PMID: 19308224
4. Niblack TL, Heinz RD, Smith GS, Donald PA. Distribution, density, and diversity of *Heterodera glycines* in Missouri. J Nematol. 1993; 25: 880–6. Available: http://www.ncbi.nlm.nih.gov/pubmed/19279857%0Ahttp://www.pubmedcentral.nih.gov/articlerender.fcgi?artid=PMC2619456 PMID: 19279857
5. Faghihi J, Ferris JM. An efficient new device to release eggs from *Heterodera glycines*. J Nematol. 2000; 32: 411–413. Available: http://journals.fcla.edu/jon/article/view/67182%5Cnhttp://www.ncbi.nlm.nih.gov/pmc/articles/PMC2620471 PMID: 19270996
6. Beeman AQ, Tylka GL. Assessing the effects of ILeVO and VOTiVO seed treatments on reproduction, hatching, motility, and root penetration of the soybean cyst nematode, *Heterodera glycines*. Plant Dis. 2018; 102: 107–113. https://doi.org/10.1094/PDIS-04-17-0585-RE PMID: 30673448
7. Hajihassani A, Dandurand L-M. An improved technique for sorting developmental stages and assessing egg viability of *Globodera pallida* using high-throughput complex object parametric analyzer and sorter. Plant Dis. 2018; 102: 2001–2008. https://doi.org/10.1094/PDIS-09-17-1428-RE PMID: 30133359
8. Tylka GL, Niblack TL, Walk TC, Harkins KR, Barnett L, Baker NK. Flow cytometric analysis and sorting of *Heterodera glycines* eggs. J Nematol. 1993; 25: 596–602. Available: http://www.ncbi.nlm.nih.gov/pubmed/19279815 PMID: 19279815
9. Jenkins WR. A rapid centrifugal-flotation technique for separating nematodes from soil. Plant Dis Report. 1964;48.
10. Deng D, Zipf A, Tilahun Y, Sharma GC, Jenkins J, Lawrence K. An improved method for the extraction of nematodes using iodixanol (OptiPrep TM). J Microbiol. 2008; 167–170.
11. Beeman AQ, Njus ZL, Pandey S, Tylka GL. Chip technologies for screening chemical and biological agents against plant-parasitic nematodes. Phytopathology. 2016; 106: 1563–1571. https://doi.org/10.1094/PHYTO-06-16-0224-R PMID: 27452899
12. Tylka GL. Acid Fuchsin Stain Preparation [Internet]. 2012. Available: https://www.plantpath.iastate.edu/tylkalab/content/acid-fuchsin-stain-preparation
13. XU Z, Cheng XE. Zebrafish tracking using convolutional neural networks. Sci Rep. Nature Publishing Group; 2017; 7: 42815. https://doi.org/10.1038/srep42815 PMID: 28211462
14. Masci J, Meier U, Cireşan D, Schmidhuber J. Stacked convolutional auto-encoders for hierarchical feature extraction. Lecture Notes in Computer Science (including subseries Lecture Notes in Artificial

Intelligence and Lecture Notes in Bioinformatics). 2011. pp. 52–59. https://doi.org/10.1007/978-3-642-21735-7_7

15. Ronneberger O, Fischer P, Brox T. U-net: Convolutional networks for biomedical image segmentation. Lect Notes Comput Sci (including Subser Lect Notes Artif Intell Lect Notes Bioinformatics). 2015; 9351: 234–241. https://doi.org/10.1007/978-3-319-24574-4_28
16. Kingma DP, Ba J. Adam: A method for stochastic optimization. Proc 12th Annu Conf Genet Evol Comput—GECCO '10. 2014; 103. https://doi.org/10.1145/1830483.1830503
17. Chollet François. Keras: The python deep learning library. keras.io. 2015. https://doi.org/10.1086/316861
18. Abadi M, Barham P, Chen J, Chen Z, Davis A, Dean J, et al. TensorFlow: A system for large-scale machine learning. 12th USENIX Symposium on Operating Systems Design and Implementation. 2016. pp. 265–283.
19. Structural analysis and shape descriptors [Internet]. 2018. Available https://docs.opencv.org/2.4/modules/imgproc/doc/structural_analysis_and_shape_descriptors.html
20. Mudanyali O, Tseng D, Oh C, Isikman SO, Sencan I, Bishara W, et al. Compact, light-weight and cost-effective microscope based on lensless incoherent holography for telemedicine applications. Lab Chip. 2010; 10: 1417. https://doi.org/10.1039/c000453g PMID: 20401422
21. Silvanmelchior. RPi_Cam_Web_Interface [Internet]. github.com; 2014. Available: https://github.com/silvanmelchior/RPi_Cam_Web_Interface
22. Isikman SO, Greenbaum A, Lee M, Bishara W, Mudanyali O. Modern trends in imaging viii: lensfree computational microscopy tools for cell and tissue imaging at the point-of-care and in low-resource settings. Anal Cell Pathol. 2012; 35: 229–247. https://doi.org/10.3233/ACP-2012-0057 PMID: 22433451
23. Dovhaliuk RY. Review of digital holography reconstruction methods. In: Angelsky O V., editor. Thirteenth International Conference on Correlation Optics. SPIE; 2018. p. 5. https://doi.org/10.1117/12.2300759
24. Bao P, Situ G, Pedrini G, Osten W. Lensless phase microscopy using phase retrieval with multiple illumination wavelengths. Appl Opt. 2012; 51: 5486–94. https://doi.org/10.1364/AO.51.005486 PMID: 22859039
25. Zuo C, Sun J, Zhang J, Hu Y, Chen Q. Lensless phase microscopy and diffraction tomography with multi-angle and multi-wavelength illuminations using a LED matrix. Opt Express. 2015; 23: 14314. https://doi.org/10.1364/OE.23.014314 PMID: 26072796
26. Van Bezooijen J. Methods and techniques for nematology. Wageningen University; 2006.
27. Akintayo A, Tylka GL, Singh AK, Ganapathysubramanian B, Singh A, Sarkar S. A deep learning framework to discern and count microscopic nematode eggs. Sci Rep. 2018; 8: 9145. https://doi.org/10.1038/s41598-018-27272-w PMID: 29904135